\documentclass[11pt]{article}

\usepackage[margin=1in]{geometry}
\usepackage{graphicx}
\usepackage{subcaption}
\usepackage{siunitx}
\usepackage{acro}
\usepackage{multicol}
\usepackage[space]{cite}
\usepackage{hyperref}

\def\beq{\begin{equation}}
\def\eeq#1{\label{#1}\end{equation}}
\def\eeqn{\end{equation}}

\def\beqa{\begin{eqnarray}}
\def\eeqa#1{\label{#1}\end{eqnarray}}
\def\eeqan{\end{eqnarray}}

\def\Dslash{\not{\hbox{\kern-4pt $D$}}}
\def\dslash{\not{\hbox{\kern-2pt $\del$}}}

\def\Title#1{\begin{center} {\Large {\bf #1} } \end{center}}
\def\Author#1{\begin{center} {\normalsize {\sc #1} } \end{center}}
\def\Institution#1{\begin{multicols}{2}\begin{center} {\footnotesize {\it #1} } \end{center}\end{multicols}}
\def\Abstract#1{\noindent {\normalsize {\bf Abstract:} {\normalfont #1}}}
\def\Conference{\vspace{4mm}\begin{raggedright} {\normalsize {\it Talk presented at the 2019 Meeting of the Division of Particles and Fields of the American Physical Society (DPF2019), July 29--August 2, 2019, Northeastern University, Boston, C1907293.} } \end{raggedright}\vspace{4mm}}
\def\Collaboration#1{\begin{center} {\large {\sc #1} } \end{center}}

\newcommand{\fref}[1]{Figure~\ref{#1}}

\newcommand{\Fref}[1]{Figure~\ref{#1}}

\newcommand{\rcite}[1]{Ref.~\cite{#1}}
\newcommand{\lfrac}[2]{#1/#2}
\newcommand{\auth}[3][,]{#3#1$^{\hyperlink{aff:#2}{#2}}$}
\newcommand{\aff}[2]{\hypertarget{aff:#1}{$^{#1}$}#2\\}
\newcommand{\email}[1]{\href{mailto:#1}{#1}}
\newcommand{\indentp}{\hspace*{\parindent}}

\newcommand{\FWHM}{\ensuremath{\text{FWHM}}}
\newcommand{\MP}{\ensuremath{\text{MP}}}
\newcommand{\orderof}[1]{\ensuremath{\mathcal{O}\left(#1\right)}}

\DeclareAcronym{CVD}{long = chemical vapor deposition}
\DeclareAcronym{sCVD}{short = scCVD, long = single-crystalline \acs{CVD}, short-indefinite = an}
\DeclareAcronym{pCVD}{long = polycrystalline \acs{CVD}}

\DeclareAcronym{TRIGA}{long = {Training, Research, Isotopes, General Atomics}, first-style = reversed}
\DeclareAcronym{JSI}{long = Jožef Stefan Institute, long-pre = {the }}

\DeclareAcronym{DUT}{long = device under test, long-plural-form = devices under test}
\DeclareAcronym{ROC}{long = readout chip}

\DeclareAcronym{FWHM}{long = full width at half maximum, short-indefinite = an}
\DeclareAcronym{MP}{long = most probable, short-indefinite = an}

\DeclareAcronym{CERN}{short = CERN, long = European Organization for Nuclear Research, long-pre = {the }, first-style = short}
\DeclareAcronym{LHC}{long = Large Hadron Collider}
\DeclareAcronym{HLLHC}{short = HL-LHC, long = High Luminosity \acs{LHC}}
\DeclareAcronym{SPS}{long = {Super Proton Synchrotron}}
\DeclareAcronym{PSI}{long = Paul Scherrer Institut, long-pre = {the }}
\DeclareAcronym{HIPA}{long = {High Intensity Proton Accelerator}}

\newlength{\figwidth}
\hypersetup{
	hidelinks,
	bookmarksnumbered,
	pdftitle  = {Recent Results from Polycrystalline CVD Diamond Detectors},
	pdfauthor = {L. Bäni et al. (RD42 collaboration)}
}

\DeclareSIUnit \clight {\text{\ensuremath{c}}}
\DeclareSIUnit \p      {\text{\ensuremath{p}}}
\DeclareSIUnit \n      {\text{\ensuremath{n}}}
\DeclareSIUnit \e      {\text{\ensuremath{e}}}
\DeclareSIUnit \sqcm   {\square\cm}
\DeclareSIUnit \sqmm   {\square\mm}

\begin{document}

%
%

\Title{Recent Results from Polycrystalline CVD Diamond Detectors}

\Collaboration{The RD42 Collaboration}

\renewcommand{\thefootnote}{\fnsymbol{footnote}}
\Author{%
\auth{24}{L.~B\"ani}$^{,}$\footnote{Speaker, Email:~\email{lukas.baeni@cern.ch}}
\auth {3}{A.~Alexopoulos}
\auth{20}{M.~Artuso}
\auth{24}{F.~Bachmair}
\auth {3}{M.~Bartosik}
\auth{23}{H.~Beck}
\auth {2}{V.~Bellini}
\auth{12}{V.~Belyaev}
\auth{19}{B.~Bentele}
\auth{27}{A.~Bes
% ...
}
}
\Institution{
\aff {1}{Universit\"at Bonn, Bonn, Germany}
\aff {2}{INFN/University of Catania, Catania, Italy}
\aff {3}{CERN, Geneva, Switzerland}
\aff {4}{INFN/University of Florence, Florence, Italy}
\aff {5}{GSI, Darmstadt, Germany}
\aff {6}{Ioffe Institute, St.~Petersburg, Russia}
\aff {7}{IPHC, Strasbourg, France}
\aff {8}{ITEP, Moscow, Russia}
\aff {9}{Jo\v{z}ef Stefan Institute, Ljubljana, Slovenia}
\aff{10}{Universit\"at Karlsruhe, Karlsruhe, Germany}
\aff{11}{CEA-LIST Technologies Avanc\'ees, Saclay, France}
\aff{12}{MEPHI Institute, Moscow, Russia}
\aff{13}{The Ohio State University, Columbus, OH, USA}
\aff{14}{Rutgers University, Piscataway, NJ, USA}
\aff{15}{University of Torino, Torino, Italy}
\aff{16}{University of Toronto, Toronto, ON, Canada}
\aff{17}{University of Bristol, Bristol, UK}
\aff{18}{Czech Technical University, Prague, Czech Republic}
\aff{19}{University of Colorado, Boulder, CO, USA}
\aff{20}{Syracuse University, Syracuse, NY, USA}
\aff{21}{University of New Mexico, Albuquerque, NM, USA}
\aff{22}{University of Manchester, Manchester, UK}
\aff{23}{Universit\"at G\"ottingen, G\"ottingen, Germany}
\aff{24}{ETH Z\"urich, Z\"urich, Switzerland}
\aff{25}{University of Tennessee, Knoxville, TN, USA}
\aff{26}{INFN-Lecce, Lecce, Italy}
\aff{27}{LPSC-Grenoble, Grenoble, France}
\aff{28}{INFN-Perugia, Perugia, Italy}
\aff{29}{California State University, Sacramento, CA, USA}
\aff{30}{University of Bergen, Bergen, Norway}
}

\renewcommand{\thefootnote}{\arabic{footnote}}

\Abstract{%
	Diamond is a material in use at many nuclear and high energy facilities due to its inherent radiation tolerance and ease of use.
We have characterized detectors based on chemical vapor deposition (CVD) diamond before and after proton irradiation.
We present preliminary results of the spatial resolution of unirradiated and irradiated CVD diamond strip sensors.
In addition, we measured the pulse height versus particle rate of unirradiated and irradiated polycrystalline CVD (pCVD) diamond pad detectors up to a particle flux of \SI{20}{\MHz\per\square\cm} and a fluence up to \SI{4e15}{\n\per\square\cm}.

}

\Conference

%
%

\section{Introduction}
\indentp
As high energy experiments are upgraded to operate at higher energies and larger intensities, more radiation tolerant detector concepts are necessary to resist the increased radiation dose.
Due to its large displacement energy~\cite{Koike_1992}, sensors based on diamond feature an inherent radiation tolerance enabling their use in nuclear and high energy physics experiments~\cite{Edwards_2004, Dong_2008, Cindro_2008, Bell_2010}.
During the last two decades, the RD42 collaboration has developed and characterized \ac{CVD} diamond as a sensor material for high radiation experiments~\cite{CERN_DRDC_94-21, CERN-LHCC-2018-015}.

Previous to this work, the RD42 collaboration has measured the radiation tolerance of \ac{CVD} diamond against protons and neutrons of various energies~\cite{Baeni_2019, Baeni_LP_2019} and studied the pulse height uniformity of \ac{CVD} diamond-based detectors as a function of particle fluence~\cite{Baeni_2019}.
In this article, we present preliminary measurements of the spatial resolution of \ac{CVD} diamond strip detectors as a function of particle fluence.

Another essential characteristic of position sensitive detectors for future high energy experiments is their capability to operate at high particle rates, up to \orderof{\si{\GHz\per\square\cm}}.
Previously, the RD42 collaboration has conducted a series of beam tests to characterize the pulse height dependence in diamond detectors on particle flux~\cite{Wallny_PIXEL_2014, Wallny_ICHEP_2016}.
No dependence of the signal response of \ac{pCVD} diamond pad detectors, which were irradiated up to a neutron fluence of \SI{5e14}{\n\per\square\cm}, on the particle flux up to \SI{10}{\MHz\per\square\cm} was observed~\cite{Wallny_ICHEP_2016}.
This study was extended and \ac{pCVD} diamond samples were irradiated up to a larger neutron fluence~\cite{Reichmann_VCI_2019}.
Preliminary results of this study up to a neutron fluence of \SI{4e15}{\n\per\square\cm} are presented in this article.

Over the last four years, the RD42 collaboration has developed and characterized \ac{CVD} diamond detectors with a 3D electrode geometry in order to enhance the radiation tolerance of diamond even further~\cite{Bachmair_2015, Kagan_Vertex_2016}.
Recent developments in the fabrication process have yielded improved devices.
The new devices with a pixel readout were recently tested~\cite{Reichmann_VCI_2019}.

\section{Radiation Tolerance}
\indentp
The RD42 collaboration has recently published~\cite{Baeni_2019} the results of a series of \ac{sCVD} and \ac{pCVD} diamonds irradiated with \SI{800}{\MeV} and \SI{24}{\GeV} protons.
For these results strip detector devices were fabricated on the diamond material and characterized in beam tests.
The shape of the signal response distribution and the material uniformity as a function of particle fluence were measured~\cite{Baeni_2019}.

\begin{figure*}[bt]
	\centering
	\begin{subfigure}[b]{0.5\linewidth}
		\centering
		\includegraphics[trim = {41.504pt 0pt 8.33217pt 0pt}]{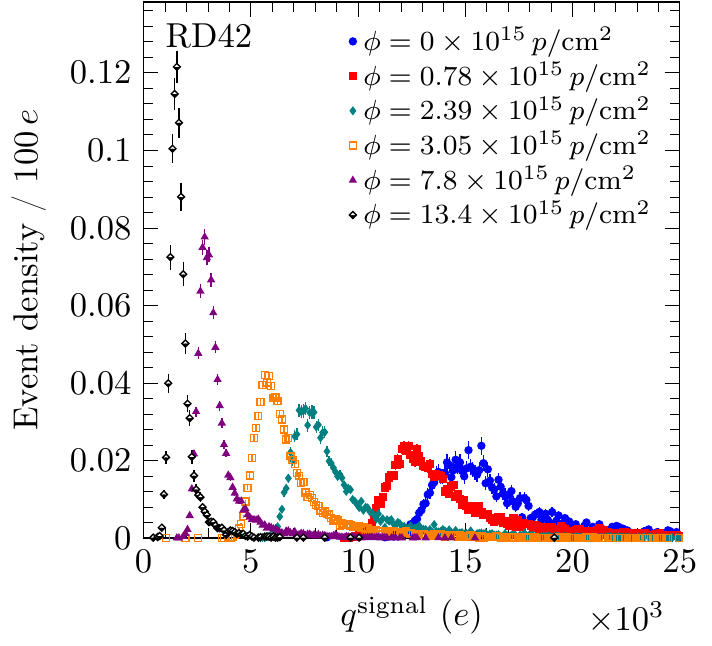}
		\caption{}
		\label{fig:PH-FluenceEvolution-neg}
	\end{subfigure}%
	\begin{subfigure}[b]{0.5\linewidth}
		\centering
		\includegraphics[trim = {41.504pt 0pt 8.33217pt 0pt}]{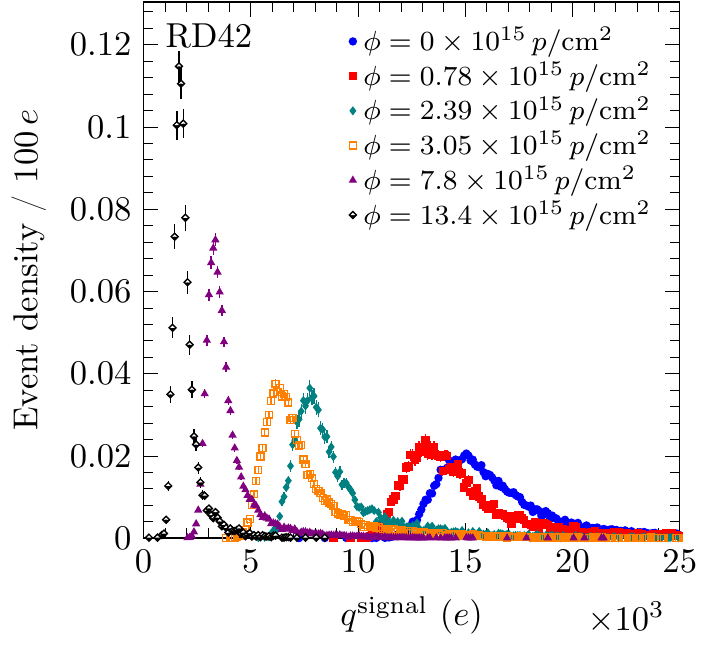}
		\caption{}
		\label{fig:PH-FluenceEvolution-pos}
	\end{subfigure}%
	\caption{Pulse height distribution before and after various \protect\SI{800}{\MeV} proton fluences at an electric bias field of (\protect\subref{fig:PH-FluenceEvolution-neg})~\protect\SI{-2}{\V\per\um} and (\protect\subref{fig:PH-FluenceEvolution-pos})~\protect\SI[retain-explicit-plus]{+2}{\V\per\um}~\protect\cite{Baeni_2019}.}
	\label{fig:PH-FluenceEvolution}
\end{figure*}

For the preliminary measurements presented here, the general aspects of the published result were followed.
Strip detectors with \SI{25}{\um} strips and a \SI{50}{\um} pitch width were manufactured on each diamond.
The devices were characterized in a beam of \SI{120}{\GeV\per\clight} hadrons at a secondary beam line of the \ac{SPS} at \ac{CERN}.
To reconstruct particle trajectories and measure the spatial resolution of the \ac{DUT} a high precision telescope~\cite{Colledani_1996} was used.
After a track selection based on the quality of the fitted track, the hit position in the plane of the \ac{DUT} was predicted with a precision of roughly \SI{1.3}{\um}~\cite{Baeni_2019}.
The signal response at the predicted position of the \ac{DUT} was measured independent of the \ac{DUT} information from the five highest adjacent strips within ten strips around the predicted track position.
\Fref{fig:PH-FluenceEvolution} shows the signal response distribution of \iac{sCVD} diamond sensor before and after various \SI{800}{\MeV} proton irradiations.
With increasing particle fluence the mean of the distribution decreases and the width becomes narrower.
To compare the shape of the signal response distribution as function of particle fluence between \ac{sCVD} and \ac{pCVD} diamond material, the \ac{FWHM} normalized by the \ac{MP} pulse height was studied.
The ratio \lfrac{\FWHM}{\MP} was observed to decrease with particle fluence for \ac{pCVD} diamond material, while for \ac{sCVD} diamond material a smaller initial ratio was measured and no dependence on the particle fluence was observed~\cite{Baeni_2019}.
The ratio \lfrac{\FWHM}{\MP} is related to the uniformity of the material, indicating \ac{pCVD} diamond becomes more uniform with particle fluence~\cite{Baeni_2019}.
Since charge sharing and thus the spatial resolution should be connected to the uniformity of the material, the two-strip spatial resolution was studied as a function of particle fluence.

In order to determine the spatial resolution, a two-strip cluster algorithm was used in the telescope planes to determine the predicted track location in the diamond detector under test.
The track position in the \ac{DUT} was determined independently of the predicted track location.
To accomplish this, in every event, the strip in the \ac{DUT} with the largest pulse height and the adjacent strip with the next largest pulse height were used in a two-strip $\eta$~algorithm to determine the \ac{DUT} hit position.
To avoid inefficient events due to extra telescope hits leading to mis-tracking, multiple scattering leading to incorrect \ac{DUT} position, or noise in the \ac{DUT}, the strip with the largest signal was required to be larger than five standard deviations above the noise on that strip.
This represented a minimum signal size of approximately \SI{400}{\e}.
Using this requirement, all \acp{DUT} were greater than \SI{99}{\percent} efficient.
In addition, to check that the hit strips in the \ac{DUT} were related to the tracks in the telescope, the digital residual, defined as the distance of the center of the strip with the highest signal to the predicted hit position, was measured.
The digital residual distribution is expected to be uniformly distributed with \ac{FWHM} equal to the strip pitch.
Verifying proper functionality of the \ac{DUT}, the \ac{FWHM} of the digital residual distribution of all devices was measured to fall within \SI{49.7 +- 0.5}{\um}.
Next, in order to get the best reconstruction of the hit position in the \ac{DUT} a two-strip $\eta$~algorithm from the charge deposition in the two adjacent strips with the highest signals was used~\cite{Belau_1983}.
This algorithm assumes that the efficiency is near one which was the case for these detectors.
The spatial resolution was measured to be the standard deviation of the full residual spectrum of the distance between predicted hit position of the telescope and measured hit position with the \ac{DUT}.
Details on the analysis procedure may be found in \rcite{Bachmair_2016, Baeni_2017}.

For unirradiated \ac{sCVD} diamond a spatial resolution of \SI{7.6 +- 1.0}{\um} was observed while for unirradiated \ac{pCVD} diamond a spatial resolution of \SI{15.0 +- 1.5}{\um} was measured.
One possibility for this difference may be due to grain boundaries allowing the charge to move laterally.
After a \SI{24}{\GeV} proton fluence of \SI{18.0e15}{\p\per\square\cm} the spatial resolution of \iac{pCVD} diamond device was found to be \SI{7.6 +- 0.8}{\um}.
After an \SI{800}{\MeV} proton fluence of \SI{12.6e15}{\p\per\square\cm} (equivalent to a \SI{24}{\GeV} proton fluence of \SI{21.1e15}{\p\per\square\cm}~\cite{Baeni_LP_2019}) a spatial resolution of \SI{8.6 +- 0.9}{\um} was measured with \iac{pCVD} diamond device.
The spatial resolution of \ac{sCVD} diamond was observed to be roughly flat as a function of particle fluence while the spatial resolution of \ac{pCVD} diamond devices improves to very close to the \ac{sCVD} diamond result consistent with the uniformity change~\cite{Baeni_2019}.

\section{Signal Dependence on Particle Flux}
\label{sec:rate_study}
\indentp
To characterize the signal response of diamond sensors as a function of particle flux, devices were tested in the $\pi$M1 beam line at the \ac{HIPA} at \ac{PSI}.
The particle flux of this beam line is controllable from \SI{1}{\kHz\per\square\cm} to \SI{20}{\MHz\per\square\cm}.
For these studies, the beam line was tuned to a \SI{260}{\MeV\per\clight} $\pi^{+}$ beam.

The diamond samples were irradiated with fast reactor neutrons at the TRIGA reactor of \ac{JSI}~\cite{Snoj_2011}.
Before and after each irradiation step, a single pad detector was manufactured from each diamond.
A single pad electrode was metallized on both sides of the diamond and connected to a fast amplifier with low electronic noise capable of handling the high particle rates.
A detailed description of the setup and the analysis can be found in \rcite{Bachmair_2016}.

\begin{figure*}[b!]
	\centering
	\begin{subfigure}[b]{0.5\linewidth}
		\centering
		\includegraphics[trim = {35.33759pt 0pt 0.2pt 0pt}]{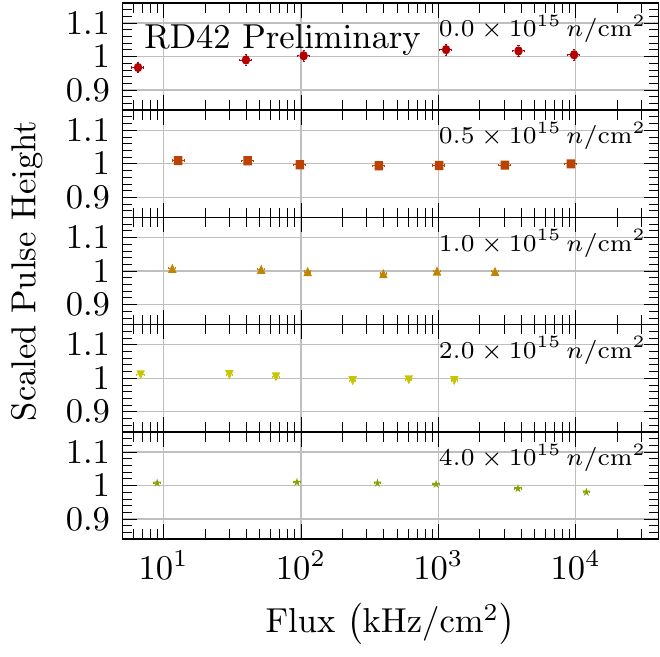}
		\caption{}
		\label{fig:rate_dependence-neg}
	\end{subfigure}%
	\begin{subfigure}[b]{0.5\linewidth}
		\centering
		\includegraphics[trim = {35.33759pt 0pt 0.2pt 0pt}]{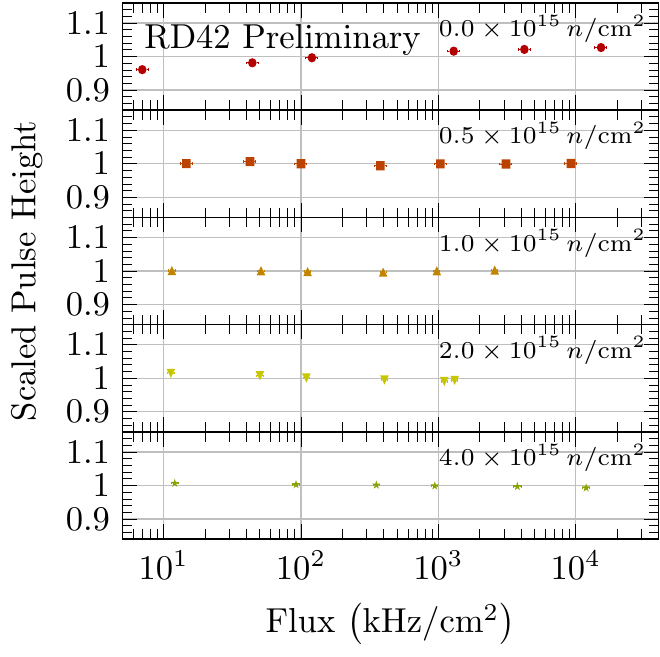}
		\caption{}
		\label{fig:rate_dependence-pos}
	\end{subfigure}%
	\caption{%
		Relative pulse height as a function of particle flux before and after fast reactor neutron irradiation at a bias voltage of
		(\protect\subref{fig:rate_dependence-neg}) \protect\SI[retain-explicit-plus]{-1000}{\V}~\protect\cite{Reichmann_VCI_2019}
		and
		(\protect\subref{fig:rate_dependence-pos}) \protect\SI[retain-explicit-plus]{+1000}{\V}.
		The pulse height was normalized to the average pulse height of each rate scan.
	}
	\label{fig:rate_dependence}
\end{figure*}

The average signal response of the device was measured at different particle rates to study the signal as a function of particle flux.
A measurement cycle included several rate scans each with negative and positive bias field applied to the detector under test.
Preliminary results of single rate scans for both negative and positive bias fields are shown in \fref{fig:rate_dependence} before and after multiple neutron irradiations.
The signal response was normalized to the average of each rate scan.
The data indicates less than a \SI{2}{\percent} variation with rate for all fluences tested.

\section{Summary}
\indentp
Preliminary results of the spatial resolution of \ac{CVD} diamond strip detectors were presented.
The spatial resolution of \ac{pCVD} diamond devices was found to improve and reach a similar value as \ac{sCVD} diamond devices.
The signal response of \ac{pCVD} diamond detectors was measured as a function of particle flux before and after several irradiation with fast reactor neutrons up to a total neutron fluence of \SI{4e15}{\n\per\square\cm}.
No rate dependence was observed at the \SI{<2}{\percent} level up to a particle flux of \SI{20}{\MHz\per\square\cm}.

\section*{Acknowledgements}
\indentp
The RD42 collaboration gratefully acknowledges the staff at CERN and PSI for test beam
time and their help in setting up the beam conditions.
The research leading to these results received funding from the European Union's
Horizon 2020 research and innovation program under grant agreement No.~654168.
This work was also partially supported by the Swiss National Science Foundation
grant 20FL20\_154216 and 20020\_134939, ETH grant 51~15-1, Swiss Government
Excellence Scholarship ESKAS No.~2015.0808, UK Science and Technology Facilities
Council grant ST/M003965/1, and the U.S.~Department of Energy through grant
DE-SC0011726.

\bibliography{references}

\end{document}